# Interaction and kinetics of $H_2$, $CO_2$, and $H_2O$ on $Ti_3C_2T_x$ MXene probed by X-ray photoelectron spectroscopy

Lars-Åke Näslund[a], Esko Kokkonen[b], and Martin Magnuson[a]

[a] *Thin Film Physics Division, Department of Physics, Chemistry, and Biology (IFM), Linköping University, SE-581 83 Linköping, Sweden.*

[b] *MAX IV Laboratory, Lund University, SE-221 00 Lund, Sweden.*

## ABSTRACT

One of the most explored MXenes is $Ti_3C_2T_x$, where $T_x$ is designated to inherently form termination species. Among many applications, $Ti_3C_2T_x$ is a promising material for energy storage, energy conversion, and $CO_2$-capturing devices. However, active sites for adsorption and surface reactions on the $Ti_3C_2T_x$-surface are still open questions to explore, which have implications for preparation methods when to obtain correct and optimized surface requirements. Here we use X-ray photoelectron spectroscopy (XPS) to study the adsorption of common gas molecules such as $H_2$, $CO_2$, and $H_2O$, which all may be present in energy storage, energy converting, and $CO_2$-capturing devices based on $Ti_3C_2T_x$. The study shows that $H_2O$, with a strong bonding to the Ti-Ti bridge-sites, can be considered as a termination species. An O and $H_2O$ terminated $Ti_3C_2T_x$-surface restricts the $CO_2$ adsorption to the Ti on-top sites and may reduce the ability to store positive ions, such as $Li^+$ and $Na^+$. On the other hand, an O and $H_2O$ terminated $Ti_3C_2T_x$-surface shows the capability to split water. The results from this study have implications for the correct selection of MXene preparations and the environment around the MXene in different implementations, such as energy storage, $CO_2$-capturing, energy conversion, gas sensing, and catalysts.

## 1. Introduction

The transition metal carbides or nitrides known as MXenes are two-dimensional (2D) materials that can be engineered for a variety of applications with fine-tuned properties [1,2]. MXenes consist of a transition metal (M), carbon or nitrogen (X), and termination species ($T_x$) forming a layered compound denoted $M_{n+1}X_nT_x$. One of the most studied MXenes is $Ti_3C_2T_x$, which consists of three Ti-monolayers and two C-monolayers stacked in an alternated sequence where the first and the last Ti-monolayers form surfaces terminated by $T_x$.

X-ray photoelectron spectroscopy (XPS) studies of $Ti_3C_2T_x$ have been performed to identify the inherently formed termination species; for example the early work of Halim et al. [3], Benchakar et al. [4], Schultz et al. [5], and Persson et al. [6]. Since XPS provides element-specific information on the electronic structure of the probed atoms, it is possible to gain information about the chemical environment around species on the $Ti_3C_2$-surface. XPS is therefore considered as a perfect tool when to investigate the complex interactions between the elements comprising MXenes. The analyses of the obtained XPS spectra presented by Halim et al., Benchakar et al., Schultz et al., and Persson et al. were performed through curve fittings, although with different approaches leading to contradictory conclusions. Halim et al. and Benchakar et al. based their analysis on the assumption that the Ti atoms in $Ti_3C_2T_x$ have different oxidation states ($Ti^{+1}$, $Ti^{+2}$, $Ti^{+3}$). Schultz et al. and Persson et al., on the other hand, based their analysis on the fact that different local environments around Ti atoms affect the Ti-orbital interaction with other elements and thus cause binding energy shifts in the components that compose the XPS spectra. A critical analysis of the four XPS studies, performed by Natu et al. [7], found that the approach of Persson et al. was the most scientific sound, except that Persson et al. had not included components in the XPS spectra assigned to the termination species OH. Natu et al. also questioned the suggestion from Persson et al. that while atomic fluorine (F) are located only at the threefold hollow face-centered cubic (fcc) site, which is formed by three surface Ti atoms above a middle monolayer Ti atom, the atomic oxygen (O) can occupy both the fcc site and a bridge site between two surface Ti atoms. Instead, Natu et al. proposed that one of the O components in the curve fitting of the O 1s XPS spectrum performed by Persson et al. should be replaced by an OH-component, but surprisingly not in the curve fitting of the Ti 2p XPS spectrum. However, the assignments of the symmetric Voigt functions used in the curve fitting of the O 1s XPS spectrum presented by Natu et al. are not consistent with the temperature-programmed XPS study performed by Persson et al. For example, the component that Natu et al. assigned to domains with only O-termination reduces significantly when F is heated away from the $Ti_3C_2T_x$-surface. In addition, the component that Natu et al. assigned to the termination species OH increases remarkably in the temperature-programmed XPS study performed in ultra-high vacuum conditions by Persson et al. although without increasing the total amount of oxygen-containing species [6]. In addition, the O 1s XPS spectrum component that Natu et al. assigned to OH was identified by Persson et al. to be atomic O on the fcc site using scanning tunneling electron microscopy (STEM) [6]. The second dispute between Natu et al. [7] and Persson et al. [6] is the designation of two features in the O 1s XPS spectrum to the same species; atomic O on the fcc site at 531.3 eV and on the bridge site at 529.9 eV. The separation by 1.9 eV may appear large, although similar binding energy separation between two surface sites is observed also in the classical example of CO adsorption on Pt(111), where CO adsorbs both on top of Pt atoms and between two Pt atoms in a bridge position giving rise to two O 1s peaks with binding energies 532.7 and 531.0 eV, respectively [8]. Hence, two different sites on the same sample surface can induce a relatively large difference in binding energy in the O 1s XPS spectrum even though the species on the surface is the same, i.e., in the present work atomic O. The molecular orbital mixing between an atom in a molecule (or a single atom) and a substrate atom depends on the available orbitals and their directions at the occupying site [9,10]. The bonding condition will therefore be dissimilar at different sites, which will affect the core hole screening of the probed atom and thus the kinetic energy of the ejected photoelectrons [11]. Hence, the XPS spectra analysis of $Ti_3C_2T_x$ presented by Persson et al. is up to this date the most reasonable. Mainly because the curve fittings of the F 1s, O 1s, Ti 2p, and C 1s XPS spectra were performed on XPS spectra obtained from a near contaminant-free $Ti_3C_2T_x$ sample with a minimum amount of impurities and that the curve fittings were based on first principles thinking, which was described comprehensively in a recent publication [12].

The method of first principles thinking requires breaking down the analysis into fundamental steps where facts are separated from assumptions and then proceed with the analysis from the ground up using those small pieces of knowledge that are self-evidently true. That is the reason why OH is not included as a termination species in the curve fitting of the XPS spectra by Persson et al. [6,12]. Even though there are numerous publications proposing OH as a termination species, there is no solid proof that the detected OH is bonded to the $Ti_3C_2T_x$-surface. Different techniques, such as Raman spectroscopy [4,13], nuclear magnetic resonance (NMR) [14], electron energy loss spectroscopy (EELS) [15,16], X-ray absorption spectroscopy (XAS) [17,18], and ultraviolet photoelectron spectroscopy (UPS) [19,20], have been employed to provide confirmation of OH as a termination species on $Ti_3C_2T_x$. Identification of specific species on a surface is, however, a challenge. It requires a well-defined sample and the sample must be nearly free from impurities and contaminations. Very few $Ti_3C_2T_x$ samples are well defined and almost all $Ti_3C_2T_x$ samples contain contaminations and impurities. For example, the parent material $Ti_3AlC_2$ is prone to oxidation [21] and $Ti_3C_2T_x$ obtained from $Ti_3AlC_2$ powders prepared in an alumina tube furnace is incorporated with small amounts of non-reacted TiC, oxidized components ($TiO_2$ and $Al_2O_3$), and components from other side reactions [19]. These impurities can pass through the MXene-forming process and become embedded in the $Ti_3C_2T_x$ sample [19] and especially $TiO_2$ is known to adsorb $H_2O$ that dissociates into OH [22-27]. The challenge of probing the termination species can be exemplified through Raman spectroscopy studies of $Ti_3C_2T_x$ performed by Sarycheva et al. [13] and Benchakar et al. [4]. Comparison with a Raman spectrum of Degussa P25 [28,29], which is a standard $TiO_2$ sample composed of anatase and rutile crystallites in about 80:20 ratio [30], shows that the Raman spectra of $Ti_3C_2T_x$ performed by Sarycheva et al. [13] and Benchakar et al. [4] have significant intensity where features from $TiO_2$ are known to appear (features at 391, 512, and 634 $cm^{-1}$ corresponds to $B_{1g}$, $A_{1g}$, and $E_g$ modes of Anatase [31] and features at 441 and 620 $cm^{-1}$ corresponds to $E_g$ and $A_{1g}$ modes of Rutile [32]). The Raman spectra comparison is presented in Figure S8. Hence, the presence of $TiO_2$ in the $Ti_3C_2T_x$ sample used in the Raman spectroscopy studies by Sarycheva et al. [13] and Benchakar et al. [4] cannot be ruled out and the structure at 289 $cm^{-1}$ that Sarycheva et al. [13] have identified as a Ti-OH feature is known to be a signature of OH on $TiO_2$ [29]. Although, the origin can as well be the $Ti_3C_2$ or residual $TiC_x$ as shown in Figure S8.

Another example where OH is detected in a $Ti_3C_2T_x$ sample is the NMR-study by Hope et al. [14]. One restriction in this study is that NMR is sensitive toward the $^{13}C$, $^{19}F$ and $^{1}H$ nuclei but not toward stable isotopes of O and Ti. They found structures in the NMR data that indicated that $^{1}H$ was bonded to O in two different compounds, most likely OH and $H_2O$, and that $^{1}H$ in OH (most likely) was located not far from $^{13}C$ and $^{19}F$. However, the NMR-study cannot verify that the presumed OH, with the detected $^{1}H$, was bonded to the $Ti_3C_2T_x$-surface as a termination species. The OH, with the detected $^{1}H$, could as well be bonded to $TiO_2$ impurities, which is invincible in the MNR-study performed by Hope et al. [14]. Since $Ti_3C_2T_x$ is a 2D-material, any oxidized regions and $TiO_2$ impurities would be in close vicinity of the $Ti_3C_2T_x$-surface and detected $^{1}H$ in OH bonded to $TiO_2$ would, thus, be located not far from $^{13}C$ and $^{19}F$. That would explain the low amount of OH estimated by Hope et al. in the NMR-study [14].

Any technique that is sensitive toward adsorption species might detect OH in the $Ti_3C_2T_x$ sample when the common impurity $TiO_2$ is present, which is known to adsorb $H_2O$ that dissociates into OH [22-27].

Even though the $Ti_3C_2T_x$ sample is free from $TiO_2$ impurities, precautions must be taken to avoid beam damage. Many techniques use photons or electrons of high energy and intensity, which might damage the sample. It has, for example, been shown that the use of high-energy electron beam irradiation in an EELS study of $Al(OH)_3$ was the origin of a feature at 533 eV in the O K-edge EELS spectrum [33]. When the energy of the electron beam was reduced the feature at 533 eV did not appear, i.e. beam damage was avoided. Further, the EELS study could not identify any feature originating from OH in the EELS data obtained from $Al(OH)_3$ and the authors therefore concluded that O K-edge EELS for the detection of OH is questionable [33].

In contrast, the combined XPS/STEM-study found no indications that OH would be a termination species [6] and a recent UPS study showed that OH was not present in a $Ti_3C_2T_x$ sample nearly free from contaminants and impurities [20]. Nevertheless, other species can be adsorbed on the $Ti_3C_2T_x$-surface, when the conditions are right, and some of them can very well become termination species after a termination species exchange process has been performed. Although, such a process is difficult to achieve [34].

While termination species are essential for the formation of the delaminated $Ti_3C_2$-layers, adsorbates are atoms and molecules that exist on the $Ti_3C_2T_x$-surface in equilibrium with the surroundings at specific conditions. This distinction is made because the termination species are remarkably stable and therefore difficult to remove or replace [34] while adsorbates exist on the surface only when the correct conditions are fulfilled. However, little is known about the interaction between $Ti_3C_2T_x$ and common gas molecules, such as $H_2$, $CO_2$, and $H_2O$.

Here, we employ high-resolution XPS to study the adsorption of $H_2$, $CO_2$, and $H_2O$ on the freestanding film of $Ti_3C_2T_x$. XPS is a suitable technique when to study the complex interactions between $H_2$, $CO_2$, and $H_2O$ and the $Ti_3C_2T_x$-surface, which affect the efficiency/quality of the $Ti_3C_2T_x$ as a catalyst, carbon capturing material, energy conversion material, gas sensing material, or any other application involving adsorption of gas molecules on the $Ti_3C_2T_x$-surface with different termination species.

## 2. Methods

### *2.1. Sample preparation*

$Ti_3C_2T_x$-flakes were produced from $Ti_3AlC_2$ powders, which were made from TiC (Alfa Aesar, 98+%), Ti (Alfa Aesar, 98+%), and Al (Alfa Aesar, 98+%) in 1:1:2 molar ratios. The formed $Ti_3AlC_2$ powder was exposed to an etching process involving 10 ml aqueous solution of 12 M HCl (Fisher, technical grade) and 2.3 M LiF (Alfa Aesar, 98+%) at 35 °C for 24 h. The produced $Ti_3C_2T_x$-flakes were contained in a deionized water suspension with a concentration of 1 mg $ml^{-1}$, which was filtered through a nanopolypropylene membrane (3501 Coated PP, 0.064 µm pore size, Celgard, USA) to make a freestanding film. The obtained $Ti_3C_2T_x$ freestanding film was thereafter stored in argon (Ar) atmosphere. Further details about the sample preparation have been presented previously [20].

The formation of $Ti_3C_2T_x$ was confirmed using X-ray diffraction (XRD) performed with an X'Pert Panalytical XRD system using Ni-filtered Cu Kα radiation in a normal Bragg−Brentano geometry. The diffractograms, presented in Figure 1(a), show no indication of $Ti_3AlC_2$ at 9.65°. Instead there is a clear peak at 7.13° as evidence for a successful $Ti_3C_2T_x$ formation. The size of the XRD peak shift through the increase in the c lattice parameter depends on the obtained spacing between the $Ti_3C_2$-layers when the Al-layers in the MAX phase are removed in the etching process [2]. The shift of the peak position from 9.65° to 7.13° corresponds to an increase of the c lattice parameter from 18.3 to 24.8 Å. The c lattice parameter increases further to 25.3 Å after sample heating, as indicated by the XRD peak shift to 6.97°. Long-range diffractograms are displayed in Figure S14.

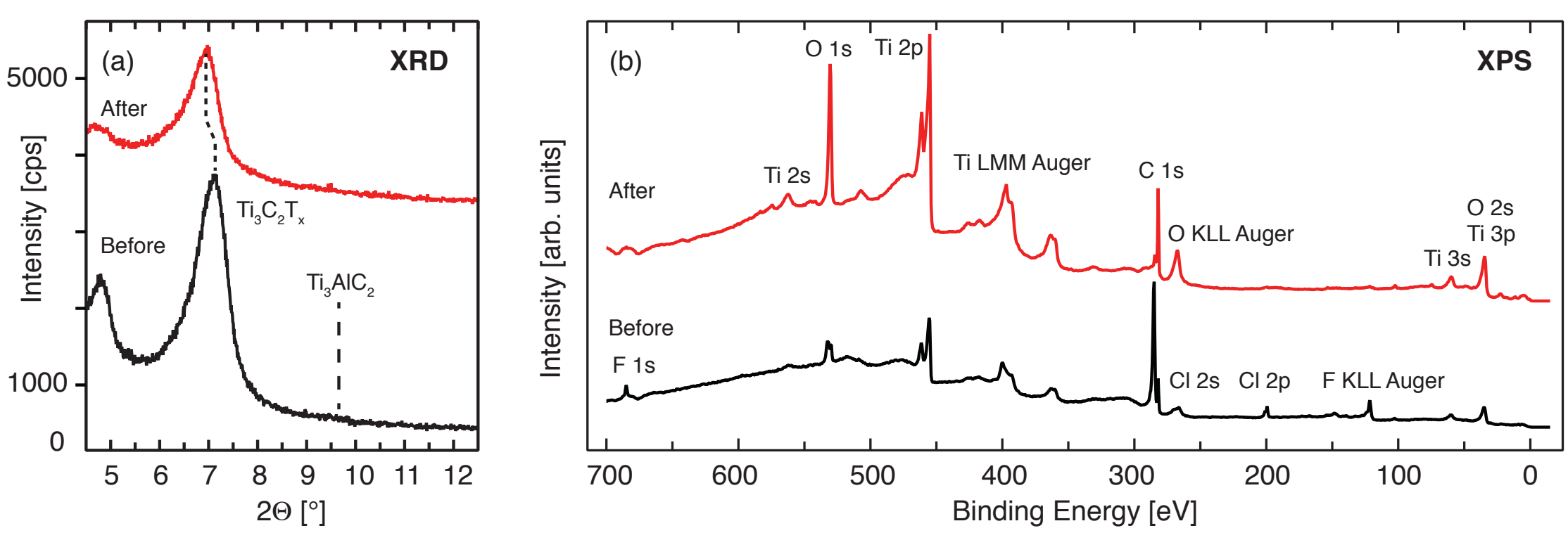

**Figure 1.** (a) X-ray diffractograms of as-prepared $Ti_3C_2T_x$ and after heat treatment at 700 °C. No intensity from $Ti_3AlC_2$ is present. (b) XPS survey spectra of as-prepared $Ti_3C_2T_x$ and after heat treatment at 700 °C. An overlayer of graphite-like carbon is removed in the heat treatment. The XPS survey spectra were recorded with a step size of 0.5 eV and pass energy of 100 eV.

The XPS survey spectra of the obtained $Ti_3C_2T_x$, presented in Figure 1(b), show that the O 1s and the Ti 2p are suppressed in the as-prepared $Ti_3C_2T_x$-sample. The reason is because of an overlayer of graphite-like carbon, as indicated by the strong C 1s. The graphite-like carbon is obtained from the etching process as a consequence from the unwanted reaction between the $Ti_3C_2$-layers and $F^-$(aq) forming $TiF_3$ and graphite-like carbon residue. The graphite-like carbon residue is lighter than the $Ti_3C_2T_x$-flakes and they are therefore separated in the centrifuging step when the freestanding film is produced [20]. The graphite-like carbon residue will thereafter descend over the formed $Ti_3C_2T_x$ freestanding film in the filtering step forming an overlayer. The detected chlorine (Cl) in the XPS survey spectrum before the heat treatment is most likely bonded to the graphite-like carbon, because it is easily removed together with the graphite-like carbon in the heat treatment, see also Figure S2(a). When mounted on the sample holder, as shown in Figure S2(b), the sample is standing with the surface normal in the horizontal plane during the heat treatments. The tension in the sample that arose in the heat treatment removed the graphite-like overlayer, probably through fragmentations, and pieces of graphite-like carbon fell down to the bottom of the vacuum chamber assisted by gravity. One benefit from the graphite-like overlayer is that the $Ti_3C_2T_x$-surface is protected against contamination when exposed to air before loading into the vacuum chamber.

### *2.2. Sample mounting and heating*

A piece of the $Ti_3C_2T_x$ freestanding film with the size of ca. 20 x 10 $mm^2$ was placed on a sample holder secured with two stainless steel clips, see Figure S2(b). A chromel/alumel thermocouple was spot welded onto the sample holder between the clips under the $Ti_3C_2T_x$ freestanding film. The sample holder system enabled monitoring of the sample temperature, both in the ultra-high vacuum (UHV) chamber and in the ambient pressure cell. In addition to the thermocouple reading, the sample temperature was also measured using a Cyclops 160L pyrometer from Land Ametek when the sample heating was performed in the UHV chamber. The sample heating in the UHV chamber was obtained through electron bombardment on the backside of the sample holder and controlled through a bias over the sample holder. In the ambient pressure cell, the sample was heated using a resistive heater located behind the sample holder. No $Ar^+$ sputtering was performed.

### *2.3. Gas exposure experiments*

The $Ti_3C_2T_x$ freestanding film was exposed to different gases in an ambient pressure gas cell. $H_2$ and $CO_2$ gases (purity 99.99% or better), obtained from Air Liquide Gas AB, were used without any extra purification. $H_2O$ vapor was obtained from deionized $H_2O$, which was cleaned with three freeze-pump-thaw cycles to remove any dissolved gases. The pressure inside the ambient pressure gas cell was monitored using a gas-independent manometer located on the outlet line of the cell. After gas exposure the ambient pressure cell was evacuated before XPS data acquisition.

### *2.4. XPS measurements*

The XPS measurements were performed using a SPECS Phoibos 150 NAP electron energy analyzer installed on the ambient pressure XPS endstation at the SPECIES beamline [35,36] at the synchrotron radiation facility MAX IV Laboratory in Lund, Sweden. The photon beam from the beamline, with a photon energy of 780 eV, impinged onto the sample at an angle of 57°, with respect to the surface normal, and provided an X-ray spot size of approximately 100 x 100 $\mu m^2$. The photoelectrons were entering the electron energy analyzer with an acceptance angle of ±22° from the sample surface normal. Since $Ti_3C_2T_x$ is a good conductor, which was confirmed through the presence of a Fermi-edge ($E_f$), there was no need for an external source of electrons to avoid sample charge-up in the photoelectron process. The $Ti_3C_2T_x$ freestanding film was instead grounded through the electron energy analyzer via the sample manipulator system, which enabled photoelectron energy calibration through alignment of the Fermi-level of the sample with the Fermi-level of the electron energy analyzer.

The Fermi-edge region was included in the recordings of each series of core-level regions (F 1s, O 1s, Ti 2p, and C 1s). Thus, the binding energy scale of each XPS spectrum was calibrated using the sample Fermi-edge, which was set to a binding energy of 0.00 ±0.02 eV [12]. An observed binding energy shift larger than ±0.02 eV can therefore be considered to be because of changes in the vicinity of the probed atom, although binding energy shifts smaller than 0.07 eV should be interpreted with caution. The core-level XPS spectra were recorded with a step size of 0.05 eV and pass energy of 50 eV, which provided an overall energy resolution better than 0.25 eV for all core-level XPS spectra.

The intensity of each XPS spectrum was normalized at the background on the low binding energy side of the main peak/peaks (i.e. each XPS spectrum was divided by its background intensity value). The background contribution in each XPS spectrum was thereafter removed using a Shirley function [12,37]. The subsequent determination of the integrated intensity of an XPS spectrum provides a dimensionless value that is proportional to the amount of probed atom in the XPS detection volume. It was, thus, possible to monitor modifications on the $Ti_3C_2T_x$ surface that were introduced by different treatments. Element quantification was, however, not possible, because suitable sensitivity factors and other system specific data that are needed for quantitative XPS analysis were not available.

*2.5. XPS spectra curve fittings*

The background contribution in each XPS spectrum was removed using a Shirley function before curve fitting of the spectrum [12,37]. The background subtractions and the curve fittings of the F 1s, O 1s, Ti 2p, and C 1s XPS spectra were performed using Igor Pro version 6.22A from WaveMetrics, which is a technical graphing and data analysis software for scientists and engineers [38]. The peak fitting procedure was based on first principles thinking as described in a previous work [12]. The XPS spectra were curve fitted using convoluted Lorentzian- and Gaussian functions, aka Voigt functions, with tails toward higher binding energies [12]. The curve fittings of the F 1s, O 1s, Ti 2p, and C 1s XPS spectra are briefly described in Figure S9-S12. The binding energy positions of the Voigt functions might differ slightly (±0.05 eV or less) from previous work [6,12,21] because of minor deviations in the kinetic energy measurements of the photoelectrons.

**3. Results and discussion**

The obtained core level XPS spectra are presented and compared in Figure S1-S6. However, thorough curve fittings of the obtained XPS spectra are needed before interpretations are possible. The curve fittings were performed according to first principles thinking as presented in a previous publication [12]. The use of different colours makes it easier to distinguish each Voigt function. The green Voigt functions represent the curve fittings of the XPS spectra obtained from the as-prepared $Ti_3C_2T_x$ sample, which confirm previous curve fittings in earlier studies [6,21]. The red Voigt functions are the most interesting Voigt functions for the present study, because they become significant after heat treatments or gas exposures. The blue Voigt function in the O 1s XPS spectra represents $Al_2O_3$ impurities, which is a residual from the etching of $Ti_3AlC_2$.

Figure 2 displays the F 1s, O 1s, Ti 2p, and C 1s XPS spectra of the as-prepared $Ti_3C_2T_x$ freestanding film. The shape of the F 1s XPS peak has, in previous works [6,12,21], been revealed to be the sum of two features located at 685.3 and 684.5 eV. Both features originate from F on the face-centered cubic (fcc) site where the latter derives from domains with only F and the former from domains with F that shares Ti-bonding with O [12]. Also the F 1s XPS spectrum in Figure 2 needs two Voigt functions with tails to fit the spectrum with binding energies 685.2 and 684.4 eV, although the Voigt function at 684.4 eV has a very low intensity indicating that domains with only F are very small in the present $Ti_3C_2T_x$ sample.

That the two Voigt functions for the F 1s XPS spectrum fit had tails is motivated by the fact that $Ti_3C_2T_x$ MXene is a good conductor because of free-moving electrons similar as in metals [21]. When a photoelectron is ejected from the probed atom the free-moving electrons will flow toward the core-hole. The photoelectron will interfere with the wave of free-moving electrons and cause harmonics. The $0^{th}$ harmonic will appear as a tail on the main peak toward higher binding energies. Figure 2(a) shows that the tail on the Voigt function at 685.2 eV fits the tail on the XPS peak very well in the binding energy region 686-690 eV. Hence, since there are no indications of additional F features in the F 1s XPS spectrum there are no reason to suspect incorporations of F-containing residuals, such as $TiO_{2-x}F_x$, $AlF_x$ and/or $Al(OF)_x$, in the sample.

Figure 2 further shows that the O 1s XPS spectrum is dominated by a feature at 529.8 eV, which originates from O bonded to two surface Ti atoms in a bridge position [6,12,20,21]. In addition, four Voigt functions with very low intensities are contributing to the curve fitting of the shoulder on the high binding energy side. Three of them are motivated by known components obtained in previous works [6,12,21], which contribute with Voigt functions at 532.4, 531.7, and 531.2 eV. The high-energy component originates from residual $Al_2O_3$ while the other two originate from O on the fcc-sites [6,39]. Previous studies have concluded that O located on the fcc-site can generate a feature at 531.7 or 531.2 eV depending on the local environment, i.e., O on the fcc-site bonds to a surface Ti that also bonds to another species or O in the fcc-site as the only species in a domain, respectively [6,12,21]. Hence, this O 1s XPS shift has similar origins as the observed F 1s XPS shift. The fourth Voigt function at 530.3 eV has not been identified previously but is motivated by the present work as it shows to be a dominating contribution to the O 1s XPS spectra in Figure 3-10.

The very small intensity of the F feature at 684.4 eV and the very small intensities of the O features at 531.7 and 531.2 eV suggest that the $Ti_3C_2T_x$ sample in this study is almost saturated by F on the fcc sites and saturated by O on the Ti-Ti bridge-sites. A combined temperature-programmed XPS and STEM study has shown that F has precedence over O regarding the fcc-site, which forces the O to the Ti-Ti bridge-sites when F is present on the $Ti_3C_2T_x$-surface [6]. Furthermore, that F 1s XPS spectrum shows that almost all F on the fcc-site are affected by the presence of O on the Ti-Ti bridge-site implies that F on the fcc-site and O on the Ti-Ti bridge-site interact with the same Ti atom. Subsequently, the O 1s XPS feature at 531.7 eV originates from O on the fcc-site that bonds to a Ti atom that also bonds to an O on the Ti-Ti bridge-site.

The curve fitting of the Ti 2p XPS spectrum in Figure 2, with the spin orbit splitting separating the Ti $2p_{3/2}$ and Ti $2p_{1/2}$ contributions, is the most challenging in the present work. For example, it is necessary to take into consideration the free-moving electrons that flow to the core holes formed in the XPS process and thus interact with the photoelectrons causing harmonics in the Ti 2p XPS spectrum [11,12]. The $0^{th}$ harmonic appears as a tail on the high binding energy side of the Ti 2p features and the 1st harmonic of the Ti $2p_{3/2}$-component contributes to the background in the Ti $2p_{1/2}$-region [11,12]. On the other hand, the absence of features in the 457.7-461.9 eV range confirms that the $Ti_3C_2T_x$ sample is free from $TiO_2$ and other titanium oxide components. The Ti 2p XPS spectrum of the as-prepared $Ti_3C_2T_x$ sample can therefore be curve fitted with three pairs of Voigt-functions with tails [6,12,21]. The components at 455.0 and 461.2 eV correspond to $Ti_3C_2T_x$ with O termination, the components at 455.9 and 462.3 eV correspond to $Ti_3C_2T_x$ with F and O termination, and the components at 456.9 and 463.3 eV correspond to $Ti_3C_2T_x$ with F termination.

The C 1s XPS spectrum in Figure 2 shows a sharp carbide peak at 282.0 eV and a relatively large graphite-like peak at 284.2 eV [40]. The carbide peak with the binding energy position at 282.0 eV is a reliable indication of a successful formation of $Ti_3C_2T_x$ [7]. The graphite-like peak originates from multilayer of graphite-like carbon flakes on top of the $Ti_3C_2T_x$ surface, see Sample preparation in the Method-section. The broad appearance is partly because of different bonding conditions for the graphite-like monolayer closest to the $Ti_3C_2T_x$ surface compared with the bulk-like vicinity for the graphite-like monolayer far away from the $Ti_3C_2T_x$ surface and the transition in between. The obtained binding energy for the photoelectrons will therefore depend on the distance from the $Ti_3C_2T_x$ surface until the bulk-like condition. Another reason for the broad appearance is the fact that the conductivity of the multilayer of graphite-like carbon is poor in the surface normal direction, which will cause charging in the graphite-like multilayer that increases with the distance from the MXene surface. Core-holes in the graphite-like carbon layers will be screened by free-moving electrons from the $Ti_3C_2T_x$ sample, although when equlibrium between the core-hole creation in the poor conducting graphite-like multilayer and electron flow from the $Ti_3C_2T_x$ sample is reached there wll be a net charge-up in the graphite-like carbon layers, which cause small shifts toward higher binding energies of the C 1s XPS peak that depends on the distance from the $Ti_3C_2T_x$ surface. Furthermore, the sample has been briefly exposed to air before loaded into the vacuum chamber and minor amounts of hydrocarbons, alcoholes, and carboxyl-groups on top of the graphite-like carbon layers can be expected. Although any minor amount of carbon contamination on top of the graphite-like carbon overlayer, they are removed together with the graphite-like carbon overlayer in the heat treatment. The hydrocarbons, alcoholes, and carboxyl-groups through desorption and the graphite-like carbon overlayer through fragmentation because of the tension in the sample that arosed in the heat treatment.

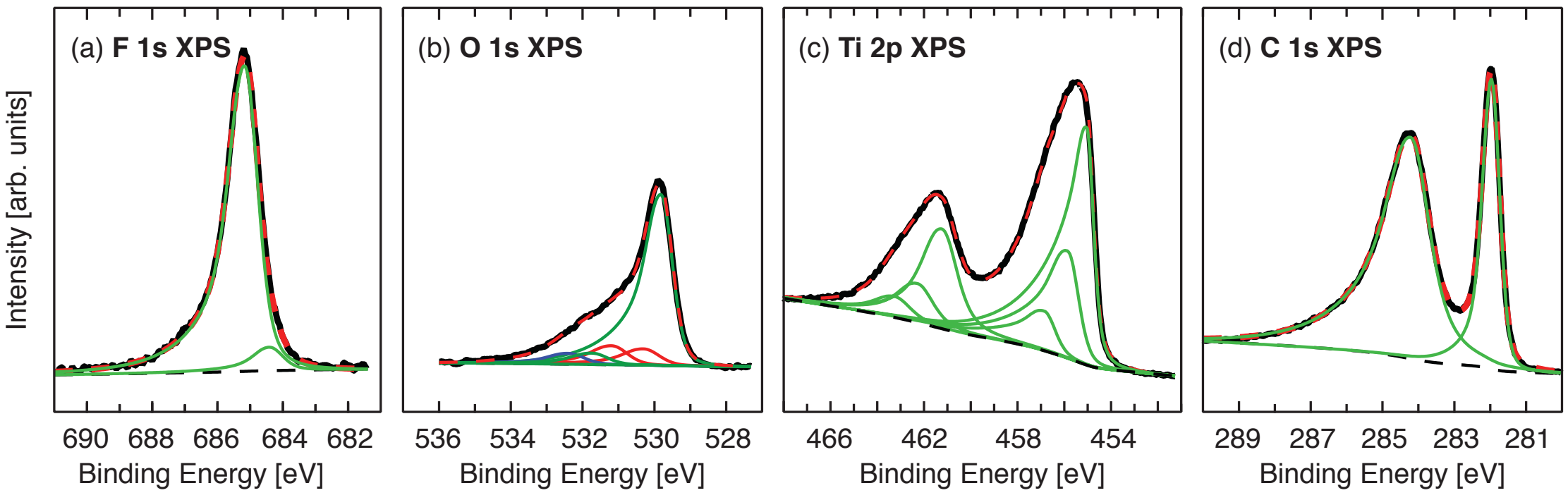

**Figure 2.** XPS spectra of the $Ti_3C_2T_x$ freestanding film for the core levels (a) F 1s, (b) O 1s, (c) Ti 2p, and (d) C 1s. The green Voigt functions represent features present in the as-prepared $Ti_3C_2T_x$ freestanding film [6,21]. The red Voigt functions at 531.2 and 530.3 eV in the O 1s XPS spectrum are defined by the curve fitting of the O 1s XPS spectrum in Figure 10. The blue Voigt function in the O 1s XPS spectrum represents $Al_2O_3$ at 532.4 eV [21,39]. The black dashed line represents the Shirley background, and the red dashed line is the total intensity of all Voigt functions in each curve fit. For further information see Supplementary material. The photon energy was 780 eV.

### *3.1. The Heat Treated $Ti_3C_2T_x$*

XPS spectra of $Ti_3C_2T_x$ after the first heat treatment at 670 °C for 30 min are presented in Figure 3. The F 1s intensity has decreased significantly and also the components at 455.9 (462.3) and 456.9 (463.3) eV in the Ti 2p XPS spectrum. Hence, the F 1s and Ti 2p XPS spectra show that the amount of F has been reduced in the heating process, as observed in previous works [5,6].

The curve fitting of the O 1s XPS spectrum in Figure 3 shows a large reduction of the peak at 529.8 eV, the vanishing of the peak at 531.7 eV, and a significant growth of the peak at 531.2 eV, which is indicative of migration of O from the Ti-Ti bridge-sites to domains with only O occupying the fcc-sites [6]. In addition there is an unexpected growth of the feature at 530.3 eV, which dominates the O 1s XPS spectrum after the heat treatment. Further, the O 1s XPS spectrum of the heat-treated $Ti_3C_2T_x$ shows an increased intensity compared with the as-prepared $Ti_3C_2T_x$. Since the integrated XPS intensity is proportional to the amount of the probed element, this finding indicates adsorption of oxygen containing species during the heat treatment. Based on the fact that the heat treatment was performed in UHV conditions, the source of the oxygen containing species must be water ($H_2O$) in the interior of the sample. Water molecules must have been intercalated between the $Ti_3C_2T_x$-flakes when the $Ti_3C_2T_x$ freestanding film was produced, which could be the reason why the flakes do not fall apart thanks to hydrogen bonding between the intercalated $H_2O$ and O on the $Ti_3C_2T_x$-surfaces [20]. During the heat treatment a significant amount of $H_2O$ molecules left the intercalated position and found the vacant Ti-Ti bridge-sites available after the migration of the O to the fcc-sites. As observed in Figure 2(b) some $H_2O$ was present on the $Ti_3C_2T_x$-surface already before the heat treatment, although only in very small amounts since most Ti-Ti bridge-sites were occupied by O that seem to have precedence. During the heat treatment the fcc-sites became vacant and therefore available for O, which migrated from the Ti-Ti bridge position making them available for the diffusing $H_2O$.

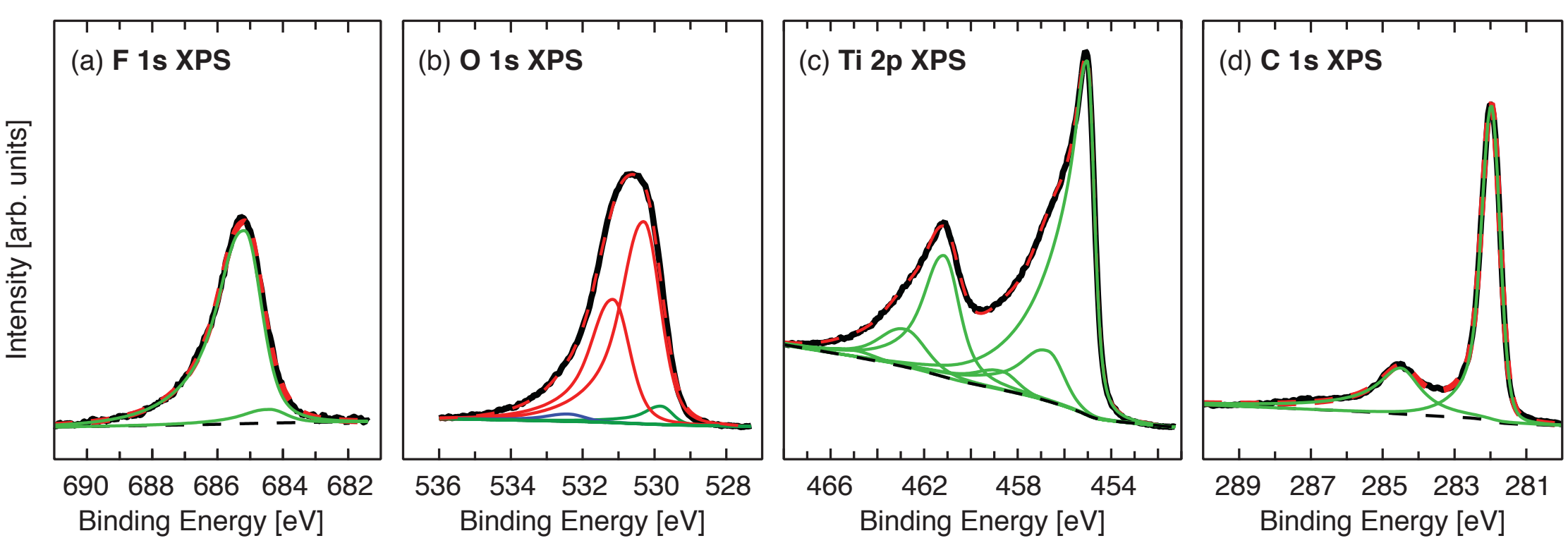

**Figure 3.** Peak fittings of XPS spectra from the $Ti_3C_2T_x$ freestanding film after the heat treatment at 670 °C for the core levels (a) F 1s, (b) O 1s, (c) Ti 2p, and (d) C 1s.

The peak at 284.2 eV in the C 1s XPS spectrum has decreased substantially as the graphite-like carbon in the sample has been removed in the heating process. The sharp carbide peak at 282.0 eV confirms the good quality of the $Ti_3C_2T_x$ freestanding film.

*3.2. $H_2$ Gas Exposure*

The sample was transferred into an ambient pressure gas cell for $H_2$ gas exposure for 3 h, where the first 2.5 h were at a sample holder temperature of 450 °C and the last 0.5 h without heating, i.e., during the cool-down of the sample. The $H_2$ gas exposure effectively removed F from the $Ti_3C_2T_x$-surface, as shown in the F 1s XPS presented in Figure S3.

Figure 4 shows that the O 1s XPS intensity has increased because of additional $H_2O$ diffusion in the heating process, from the bulk of the sample to the surface. Surprisingly, the feature at 531.2 eV, which is assigned to O on the fcc-sites in domains with only O, has decreased and a new feature at 529.8 eV has appeared. Hence, the adsorption of H has forced the O away from the fcc-sites and back into the Ti-Ti bridge-sites. The O on the Ti-Ti bridge-sites seems to be slightly affected by the presence of H and $H_2O$ and the fitted peak shifted slightly toward lower binding energies. This slight shift is an indication of the small effect the H adsorption has on the molecular orbitals available on the Ti-Ti bridge site while it has a significant effect on the Ti molecular orbitals available on the fcc-site.

The substantial effect on the Ti molecular orbitals is also observed in the Ti 2p XPS spectrum presented in Figure 4, which shows considerable large alterations as two new Ti $2p_{3/2}$ features, represented by Voigt functions with tails at 456.5 and 458.4 eV. The corresponding features in the Ti $2p_{1/2}$ XPS spectrum, i.e., at 462.6 and 464.1 eV, are not pronounced, which can be because the background intensity contribution is not always properly represented by a Shirley function [11,12] as the 1$^{st}$ harmonic from the interaction between the photoelectron and the free-flowing electrons in the $Ti_3C_2T_x$-sample coincide with the Ti $2p_{1/2}$ features (as discussed above). The Ti $2p_{3/2}$ feature at 456.5 eV (and the Ti $2p_{1/2}$ feature at 462.6 eV) is connected to the appearance of the $H_2O$ on the $Ti_3C_2T_x$ surface and is therefore considered to originate from Ti bonded to $H_2O$ on the Ti-Ti bridge-sites. The Ti $2p_{3/2}$ feature at 458.4 eV (and the Ti $2p_{1/2}$ feature at 464.1 eV) originates from the adsorption of H on the Ti on-top sites, which is 1.5 eV higher compared with the Ti $2p_{3/2}$ feature for F on the fcc-sites. The high binding energy position for Ti with H adsorbed on top is remarkable considering H has a much lower electronegativity than F (2.20 vs 3.98 by Pauling scale). However, a recent Ti 1s XANES study showed that the molecular orbitals with Ti 3d and Ti 4p characters are very sensitive toward termination species and adsorbates [18]. Hence, the interaction between the H and the molecular orbital available on the Ti on-top site causes a perturbation of the valence orbitals, both the occupied and the unoccupied, which change the core hole screening causing a reduction in the kinetic energy of the photoelectrons that are ejected from the probed Ti atoms [11]. In the Ti $2p_{3/2}$ XPS spectrum, this is manifested as a redistribution of intensity from 455.0 to 458.4 eV. Similar intensity redistribution, caused by changes in the Ti core hole screening, was observed when Al was deposited on TiN [41]. The perturbation of the occupied valence orbitals will, in addition, make the bonding between Ti and O on the fcc-sites less favorable and, thus, forcing the O back to the Ti-Ti bridge-sites.

The C 1s XPS carbide peak in Figure 4 shows a shift toward lower binding energies by -0.2 eV after the $H_2$ gas exposure, which suggests that the hydrogen atoms contribute charge to the two C layers in the $Ti_3C_2T_x$ crystal structure upon H adsorption. Hence, the binding energy position of the C 1s carbide peak, which so far has been steady at 282.0 eV for $Ti_{n+1}C_nT_x$ [7], shows a shift of the C 1s binding energy that correlates to the amount of H on the $Ti_3C_2T_x$-surface. In addition, the negative shift of the whole carbide peak infers a non-localized distribution of the extra electrons in the two C layers. The intensity of the carbide peak has remarkably been reduced after the $H_2$ exposure.

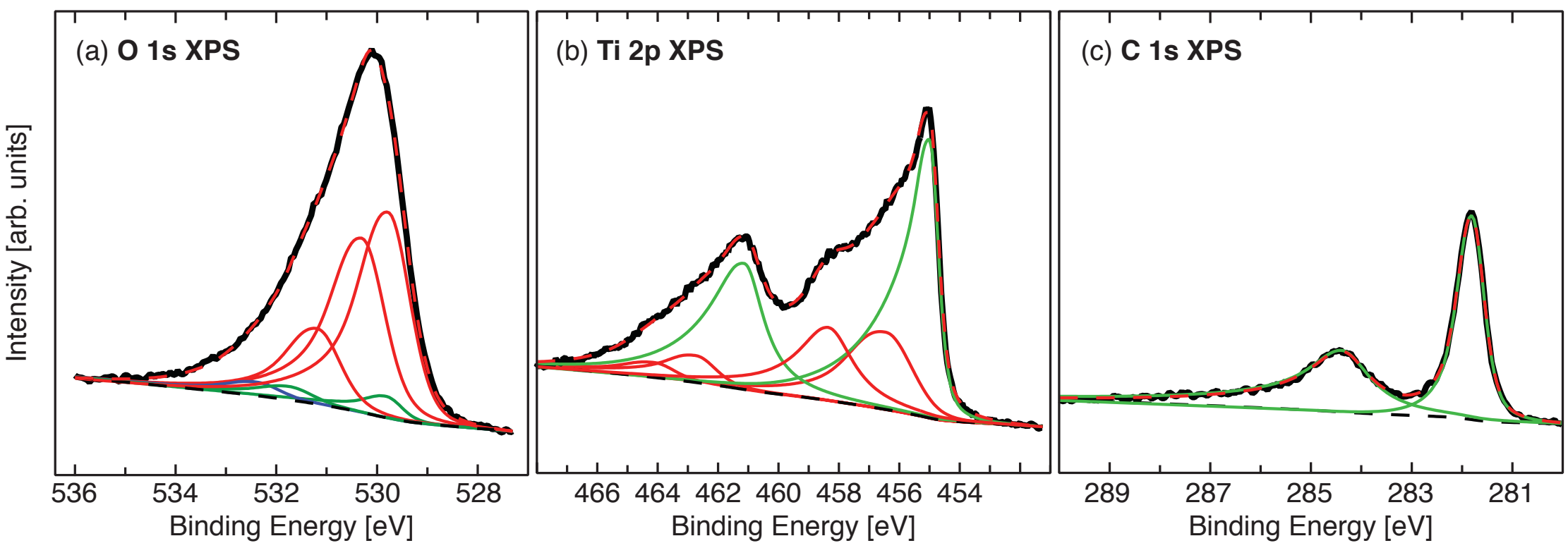

**Figure 4.** Peak fittings of XPS spectra from the $Ti_3C_2T_x$ freestanding film after the $H_2$ exposure for the core levels (a) O 1s, (b) Ti 2p, and (c) C 1s.

### *3.3. $CO_2$ Gas Exposure 1*

Without removing the adsorbed H the sample was exposed to $CO_2$. Figure 5 shows the XPS spectra obtained after the H saturated $Ti_3C_2T_x$-surface was exposed to 3.5 mbar $CO_2$ for 1.5 h at 250 °C. The XPS spectra show only minor changes and there are very small intensity variations in the peak fittings compared to the corresponding peak fitting presented in Figure 4. Although some $CO_2$ might have been adsorbed, the small differences in the XPS spectra are rather reflected by the local variations of coverage at different measurement spots. The integrated intensities for O 1s and C 1s remained the same before and after the $CO_2$ dosing, which suggests that no adsorption of $CO_2$ occurred simply because no adsorption sites were available.

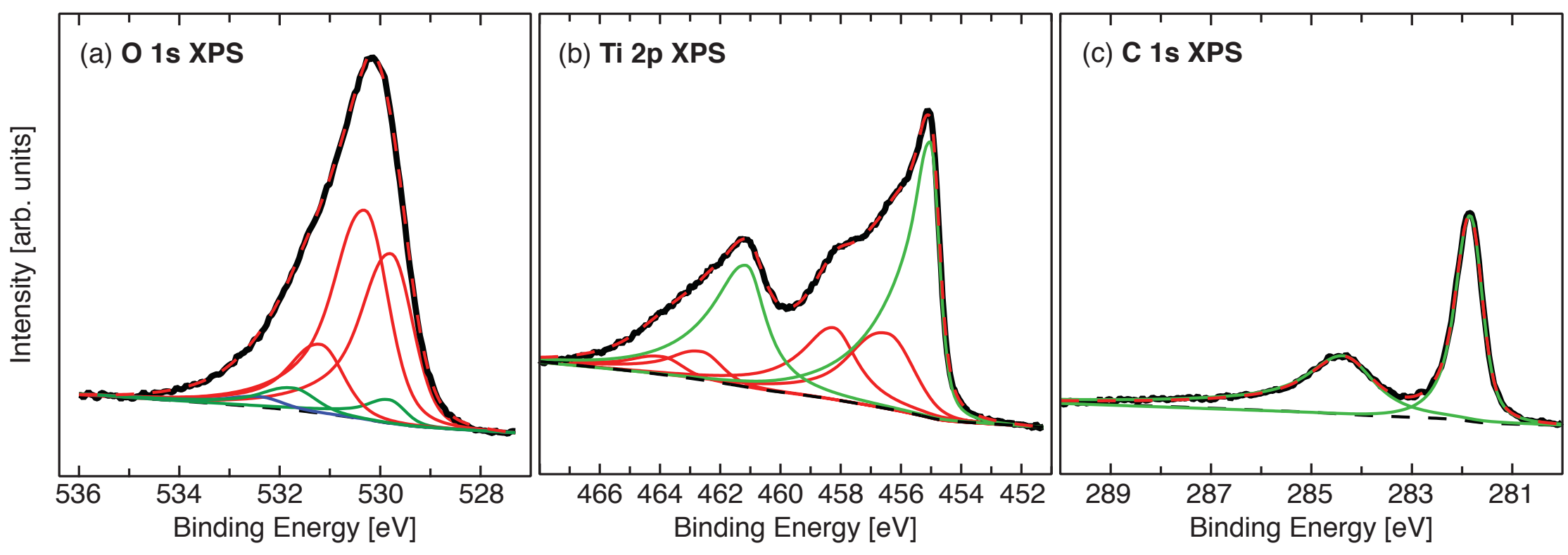

**Figure 5.** Peak fittings of XPS spectra from the $Ti_3C_2T_x$ freestanding film after the $H_2$ + $CO_2$ exposure for the core levels (a) O 1s, (b) Ti 2p, and (c) C 1s.

*3.4. $CO_2$ Gas Exposure 2*

A heat treatment at 450 °C for 35 min removed most of the adsorbed H, which is shown in Figure 6. However, the Ti 2p XPS spectrum in Figure 6 shows that there still is some additional intensity at 458.4 eV indicating that some H remained on the $Ti_3C_2T_x$-surface, which the carbide peak position at 281.9 eV in the C 1s XPS spectrum corroborates.

Nevertheless, Figure 6 shows that the O 1s XPS spectrum has different intensity redistribution after the heat treatment. The features at 531.2 and 529.8 eV assigned to O on the fcc-sites and Ti-Ti bridge-sites, respectively, show reduced intensity indicating a loss of atomic O on the $Ti_3C_2T_x$-surface. Hence, in the heating process the adsorbed H reacted with O and formed $H_2O$. In the first H addition step an OH-intermediate must be formed, which remained stable long enough for the second H to arrive and react to form $H_2O$ [42]. However, it cannot be excluded that OH-radicals were desorbed in the heating process before a second H had arrived. A possible reduction of the coverage of O-containing species on the $Ti_3C_2T_x$-surface would then allow for more diffusion of $H_2O$ from intercalation positions in the bulk of the sample up to the $Ti_3C_2T_x$-surface. The increase of $H_2O$ on the $Ti_3C_2T_x$-surface is confirmed by the dominating feature in the O 1s XPS spectrum at 530.3 eV, which is the one that originates from $H_2O$ on the Ti-Ti bridge-sites.

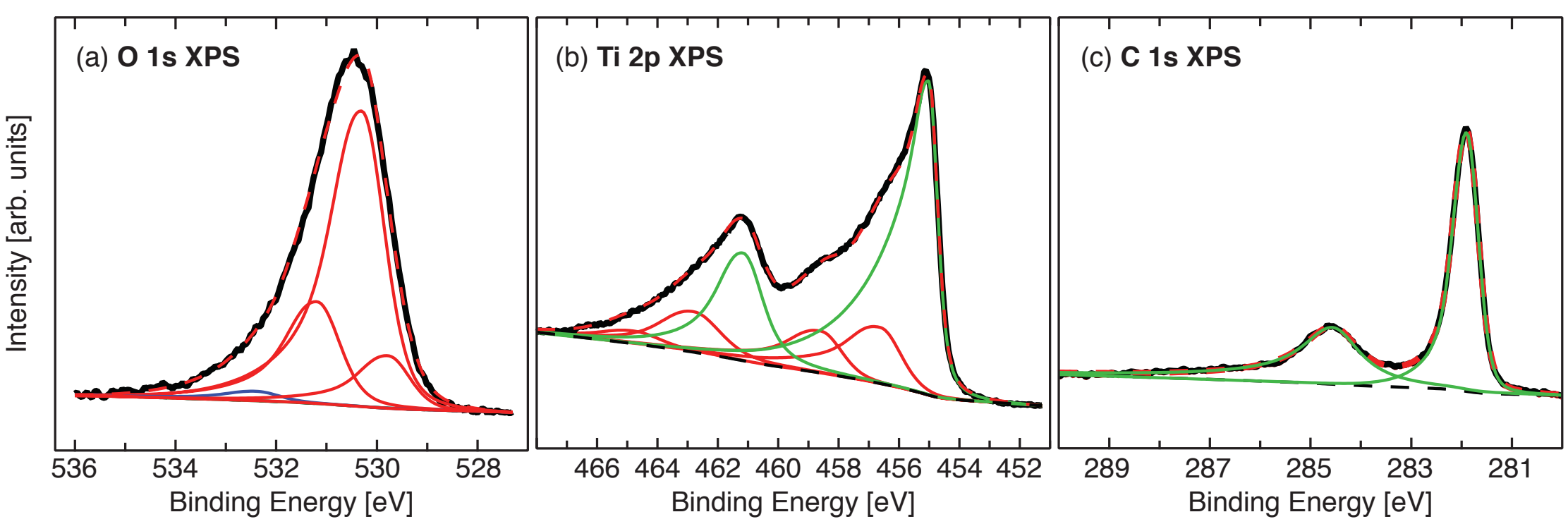

**Figure 6.** Peak fittings of XPS spectra from the $Ti_3C_2T_x$ freestanding film after the second heat treatment for the core levels (a) O 1s, (b) Ti 2p, and (c) C 1s.

With most of the H removed from the $Ti_3C_2T_x$-surface the sample was exposed to 11 mbar $CO_2$ at 450 °C for 1 h and Figure 7 shows that adsorption of $CO_2$ was thereafter possible. From a recent study of $CO_2$ adsorption on transition metal carbides we can infer that the increased integrated intensities for the O 1s XPS peak around 530.3 eV and C 1s XPS peak around 284.6 eV indicate $CO_2$ adsorption [43]. Since the increase of intensities in the O 1s and C 1s XPS spectra coincide with intensities from $H_2O$ termination on Ti-Ti bridge-site and graphite-like carbon, respectively, no additional Voigt functions were needed in the curve fitting procedure. In addition, the absence of features in the binding energy region 286-290 eV in the C 1s spectra in Figure 7 infers that there are no significant amounts of alcohol, carboxyl nor carbonate formed on the $Ti_3C_2T_x$-surface upon $CO_2$ adsorption [21,44,45]. Hence, the $CO_2$ contributes with intensity at 530.3 eV in the O 1s XPS spectrum [43], i.e., at the same binding energy position as $H_2O$ on the Ti-Ti bridge-site, and at 284.6 eV in the C 1s XPS spectrum [43], i.e. at the same binding energy position as graphite-like components. The adsorption site of $CO_2$ is, however, different compared with $H_2O$. In addition, the adsorption of $CO_2$ has forced the O away from the fcc-site and back to the Ti-Ti bridge-site as indicated by the intensity decrease at 531.2 eV and the intensity increase at 529.8 eV. This response has $CO_2$ adsorption in common with $H_2$ adsorption (see Figure 4).

Furthermore, the Ti 2p XPS spectrum in Figure 7 shows a clear feature at 458.7 eV. An oxidized $Ti_3C_2$-layer would provide a Ti $2p_{3/2}$-feature at 459.0 eV [21], which is very close to the observed Ti $2p_{3/2}$-feature after $CO_2$ adsorption. In addition would an oxidized $Ti_3C_2$-layer show an O 1s-feature at 530.6 eV [21], which is very close to the observed increase in intensity in the O 1s XPS spectrum at 530.3 eV. However, $Ti_3C_2T_x$ MXene does not oxidize easily [21]. For example, while $Ti_3AlC_2$ and TiC samples show a clear $TiO_2$-feature in the Ti 2p XPS spectra after a brief air exposure, which introduce much more $CO_2$ gas over the sample surface than in the present study, there are no indications of oxidation of an $Ti_3C_2T_x$ sample [21]. Even after removal of termination species and exposure of 3.5 mbar $CO_2$ at 100 °C for 0.5 h in a combined in situ environmental TEM/EELS study, no $TiO_2$ formation in the $Ti_3C_2$ sample was observed [46]. Lastly, exposing the $Ti_3C_2T_x$-sample to $H_2$ gave a similar feature in the Ti $2p_{3/2}$ XPS spectrum at 458.4 eV, see Figure 4, and in that condition $TiO_2$ formation can be excluded. Hence, oxidation of the $Ti_3C_2T_x$ sample when exposure to $CO_2$ can be removed from consideration.

The similarities shown with $CO_2$ and $H_2$ adsorption, and the fact that $CO_2$ does not adsorb when the $Ti_3C_2T_x$-surface is covered by H, suggest that both $CO_2$ and $H_2$ adsorbs on the Ti on-top sites at the $Ti_3C_2T_x$-surface, where the latter dissociates.

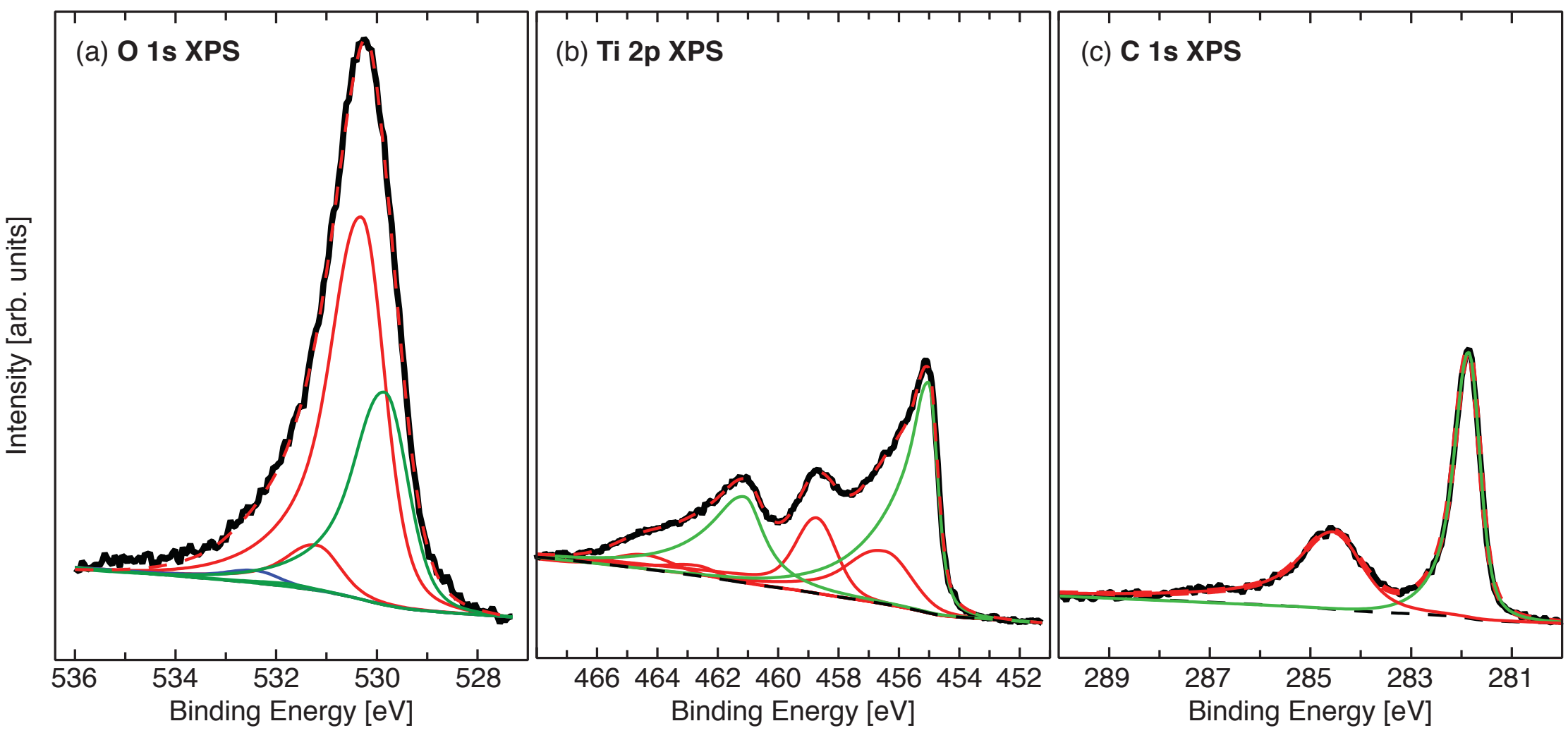

**Figure 7.** Peak fittings of XPS spectra from the $Ti_3C_2T_x$ freestanding film after the $CO_2$ exposure for the core levels (a) O 1s, (b) Ti 2p, and (c) C 1s.

A heat treatment at 650 °C desorbed the $CO_2$ from the $Ti_3C_2T_x$-surface, as shown in Figure 8, and the O could then move back to the fcc-sites. Hence, with the $CO_2$ removed from the $Ti_3C_2T_x$-surface the two dominating features in the O 1s XPS spectrum originate from the O on the fcc-sites at 531.2 eV and the $H_2O$ on the Ti-Ti bridge-sites at 530.3 eV. A small contribution from residual $Al_2O_3$ at 532.4 eV is also present [39]. The carbide peak in the C 1s XPS spectrum reappears at 282.0 eV indicating that no significant amount of H is present on the $Ti_3C_2T_x$-surface.

However, the curve fitting of the Ti 2p XPS spectrum shows Voigt functions representing Ti $2p_{3/2}$ features at 456.9 and 458.7 eV and the corresponding Voigt functions at 462.9 and 464.4 eV. The features at 456.9 and 462.9 eV are assigned to Ti bonded to $H_2O$ on a saturated $Ti_3C_2T_x$-surface. The features at 458.7 and 464.4 eV, which were attributed to the perturbation of the occupied Ti valence orbitals, suggest that some small amount of H or $CO_2$ (or perhaps CO) might still be present on the Ti on-top sites or that the $H_2$ treatment might have introduced an irreversible change in the surface structure.

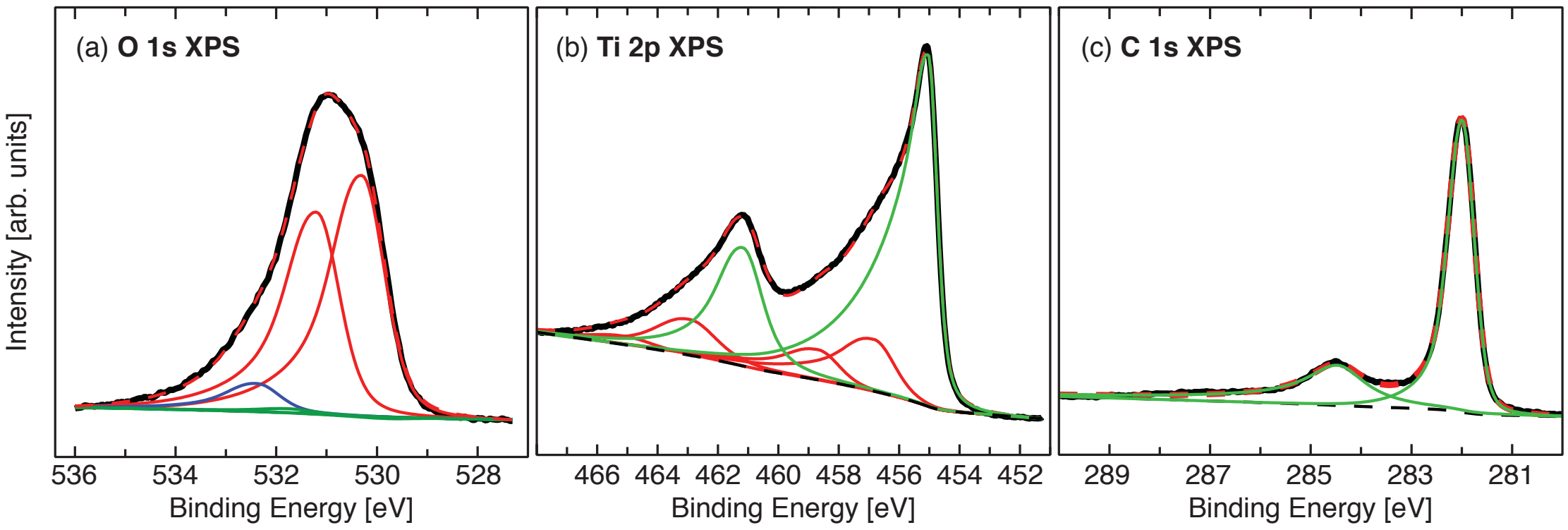

**Figure 8.** Peak fittings of XPS spectra from the $Ti_3C_2T_x$ freestanding film after the third heat treatment for the core levels (a) O 1s, (b) Ti 2p, and (c) C 1s.

### *3.5. $H_2O$ Gas Exposure*

Figure 9 shows that exposing the $Ti_3C_2T_x$-surface to $H_2O$ gas will change the shape of the O 1s XPS spectrum, although without increasing the integrated intensity. The curve fitting of the O 1s XPS spectrum displayed in Figure 9 indicates that the integrated intensity of the $H_2O$ feature at 530.3 eV is intact, which must be because the Ti-Ti bridge-sites are already saturated by $H_2O$. Instead there is a decrease of the O feature at 531.2 eV and an increase of the O feature at 531.7 eV, where the latter is O on the fcc-sites influenced by another species as shown in previous works [6,12]. Since the Ti-Ti bridge-sites are saturated by $H_2O$ there are no other options for the O except remaining at the fcc-sites, despite the perturbation of Ti molecular orbitals. In addition, there is an intensity contribution at 529.8 eV indicating small amounts of O on the Ti-Ti bridge-sites.

Furthermore, the perturbation of Ti molecular orbitals is manifested as intensity increases at 458.8 and 464.5 eV in the Ti 2p XPS spectrum. In addition there is a small negative shift of the carbide peak in the C 1s XPS spectrum, although it is less than -0.1 eV. The shift from 531.2 to 531.7 eV in the O 1s XPS spectrum, the perturbation of Ti molecular orbitals, and the negative shift of the carbide peak suggest that the adsorbed $H_2O$ has dissociated and some H remained on the Ti on-top sites, yet with O on the fcc-sites. That O persisted on the fcc-sites pushed the Ti molecular orbital perturbation features by 0.4 eV compared to when O was forced away by the H adsorption leaving the fcc-sites empty. The presence of the O on the fcc-sites is probably also the reason why the carbide shift is so small. Mainly because the contributed negative charge from the absorbed H might be constrained by the bonding between Ti and the O on the fcc-sites.

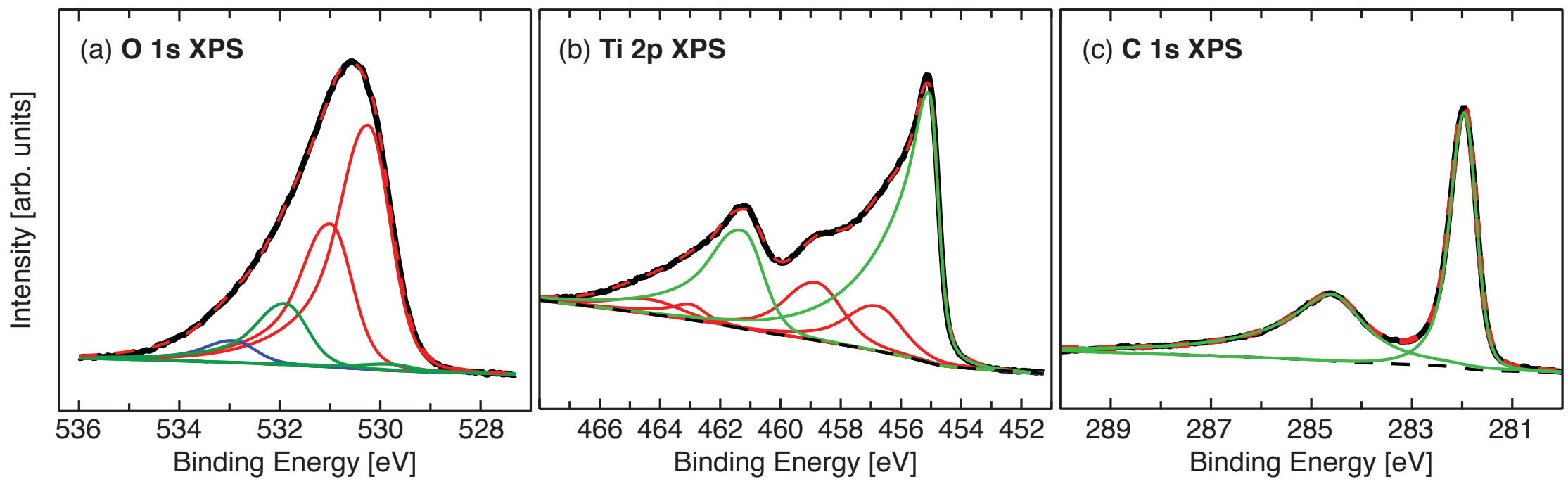

**Figure 9.** Peak fittings of XPS spectra from the $Ti_3C_2T_x$ freestanding film after the $H_2O$ exposure for the core levels (a) O 1s, (b) Ti 2p, and (c) C 1s.

Hence, the O 1s, Ti 2p, and C 1s XPS spectra in Figure 9 reveal a catalytic reaction on the $Ti_3C_2T_x$-surface where $H_2O$ adsorbs and dissociates. In addition, since the integrated intensity of the O 1s XPS spectrum did not increase upon $H_2O$ exposure, the formed O-containing species must desorb, probably as OH radicals although $O_2$ cannot be excluded.

A heat treatment at 700 °C will once again remove most of the adsorbed H and the O 1s, Ti 2p, and C 1s XPS spectra and the curve fittings presented in Figure 10 are close to identical to the corresponding spectra and curve fittings presented in Figure 8.

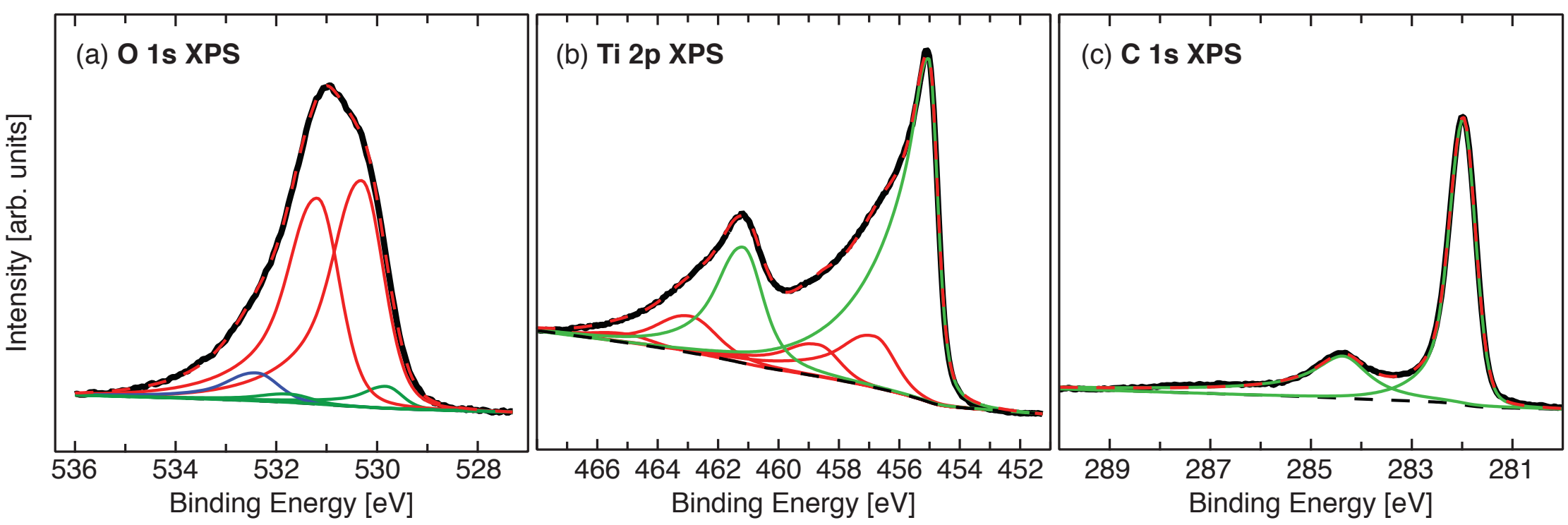

**Figure 10.** Peak fittings of XPS spectra from the $Ti_3C_2T_x$ freestanding film after the fourth heat treatment for the core levels (a) O 1s, (b) Ti 2p, and (c) C 1s.

The peak positions of the Voigt functions used in the curve fittings correspond to the binding energies of the Ti and C in the $Ti_3C_2$-layer, the terminating species F, O, and $H_2O$, and the adsorbate $CO_2$. The obtained C 1s, T 2p, O 1s, and F 1s XPS binding energies are presented in Table 1. Included in Table 1 are also the common impurities graphite-like carbon and $Al_2O_3$, which are residual components from the wet etching procedure of the parental material $Ti_3AlC_2$.

### *3.6. O 1s XPS Integrated Intensity*

Many factors, such as sample geometry, surface roughness, X-ray source, electron transmission through the spectrometer, photoelectron cross-section, etc., affect the XPS spectrum intensity [11]. The intensity of each O 1s XPS spectrum was therefore normalized at the background on the low binding energy side of the main peak, i.e. in the 527-528 eV region. The intensity of each O 1s XPS spectrum presented in Figure 2-10 is therefore proportional to the amount of O-containing species present on the $Ti_3C_2T_x$-surface after each treatment. In addition, through the curve fitting presented in this study, it is possible to monitor the relative distribution of O-containing species, O migration between different surface sites on the $Ti_3C_2T_x$ surface, and adsorption/desorption of O-containing species that was introduced by different treatments. The progressions of the integrated intensity of the components in the O 1s XPS spectra are presented in Figure 11.

**Table 1.** Binding energies for XPS features in $Ti_3C_2T_x$ and modified $Ti_3C_2T_x$ samples.

| Region | Binding Energy [eV] | Designation |
|---|---|---|
| C 1s | 281.8 | $Ti_3C_2T_x$ with adsorbed H |
| | 282.0 | $Ti_3C_2T_x$ |
| | 284.2 | multilayer graphite-like carbon |
| | 284.5 | graphite-like carbon on $Ti_3C_2T_x$ |
| | 284.6 | $CO_2$ on $Ti_3C_2T_x$ |
| Ti $2p_{3/2}$ | 455.0 | $Ti_3C_2T_x$ with O termination |
| | 455.9 | $Ti_3C_2T_x$ with F and O termination |
| | 456.5 | $Ti_3C_2T_x$ with $H_2O$ termination |
| | 456.8 | $Ti_3C_2T_x$ with $H_2O$ and O termination |
| | 456.9 | $Ti_3C_2T_x$ with F termination |
| | 458.4 | $Ti_3C_2T_x$ with adsorbed H on Ti on-top site |
| | 458.7 | $Ti_3C_2T_x$ with adsorbed $CO_2$ on Ti on-top site |
| Ti $2p_{1/2}$ | 461.2 | $Ti_3C_2T_x$ with O termination |
| | 462.3 | $Ti_3C_2T_x$ with F and O termination |
| | 462.6 | $Ti_3C_2T_x$ with $H_2O$ termination |
| | 462.9 | $Ti_3C_2T_x$ with $H_2O$ and O termination |
| | 463.3 | $Ti_3C_2T_x$ with F termination |
| | 464.1 | $Ti_3C_2T_x$ with adsorbed H on Ti on-top site |
| | 464.4 | $Ti_3C_2T_x$ with adsorbed $CO_2$ on Ti on-top site |
| O 1s | 529.8 | O termination on Ti-Ti bridge-site |
| | 530.3 | $H_2O$ termination on Ti-Ti bridge-site |
| | 530.3 | $CO_2$ adsorbed on Ti on-top site |
| | 531.2 | O termination on fcc-sites in domains with only O |
| | 531.7 | O termination on fcc-sites in domains with F |
| | 532.4 | Residual $Al_2O_3$ |
| F 1s | 684.4 | F termination on fcc-sites in domains with only F |
| | 685.2 | F termination on fcc-sites in domains with O |

At the first heat treatment the O on the Ti-Ti bridge-sites migrated to the fcc-sites and in addition the Ti-Ti bridge-sites became occupied by $H_2O$. It appears as the total amount of O reduced slightly during the heat treatment, although it can be because of local variations of coverage on the $Ti_3C_2T_x$-surface; different positions for the measuring spot were used at every experiment step.

When the $Ti_3C_2T_x$ sample was exposed to $H_2$ at 450 °C the amounts of O on the fcc-sites and $H_2O$ on the Ti-Ti bridge-sites were slightly reduced while the amount of O on the Ti-Ti bridge-sites increased significantly. The presence of H on the Ti on-top sites forced some O on the fcc-sites to the Ti-Ti bridge-sites. However, the main reason for the increased amount of O on the Ti-Ti bridge-sites must be because of $H_2O$ diffusion from the bulk followed by dissociation at the $Ti_3C_2T_x$-surface.

The second heat treatment removed a large amount of O from the Ti-Ti bridge-sites, which indicates that a significant amount of H reacted with O before desorption. However, the temperature at this heat treatment was only 450 °C and the Ti 2p and C 1s spectra in Figure 6 indicate that this heat treatment did not provide enough energy to remove all H. Although a possible recombination process forming $H_2$ that desorbs, Figure 11 suggests that most H reacted with O forming OH and $H_2O$ that desorbed from the $Ti_3C_2T_x$-surface. With the low amount of O on the $Ti_3C_2T_x$-surface the amount of $H_2O$ on the Ti-Ti bridge-sites reached the highest number.

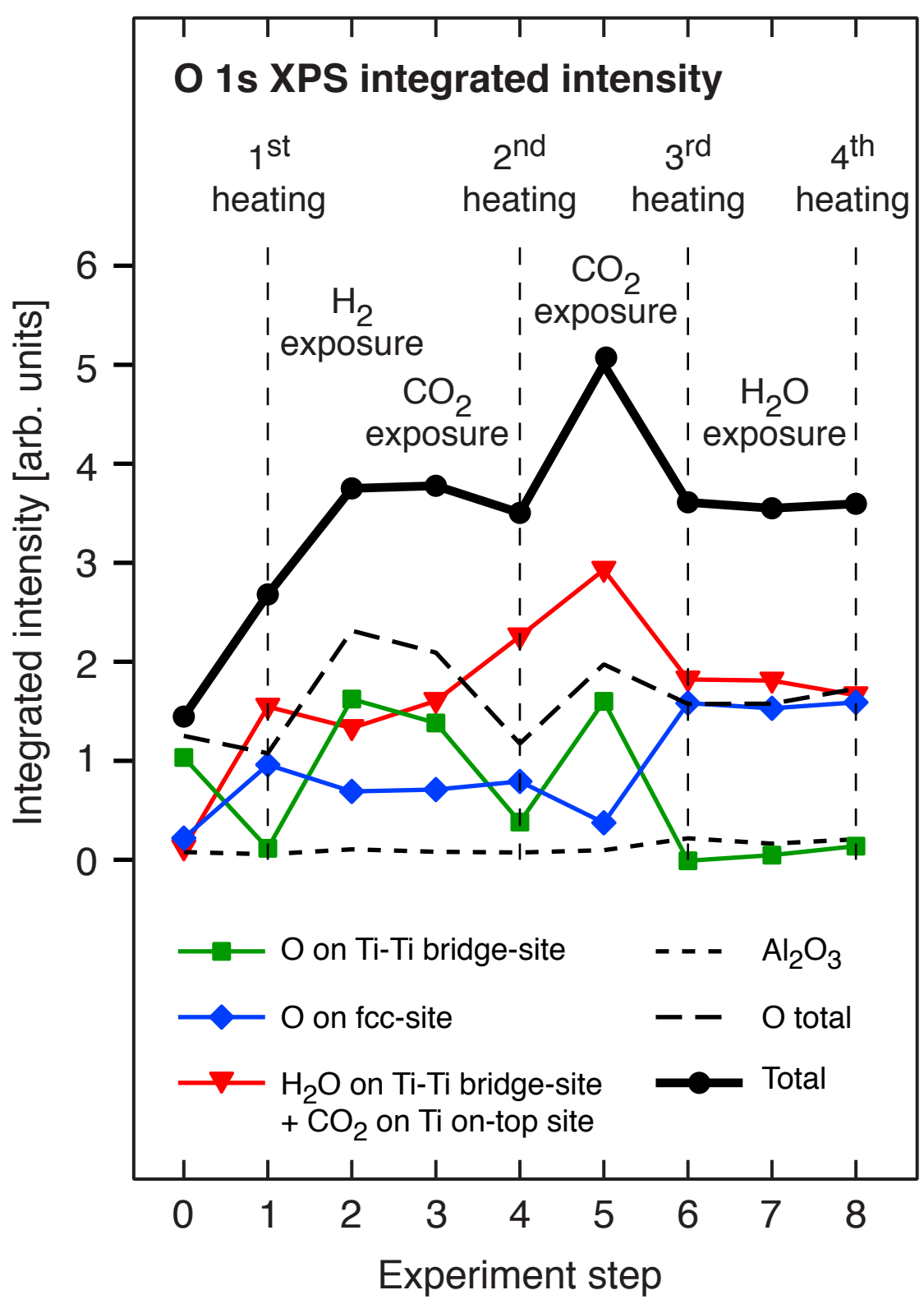

**Figure 11.** The O 1s XPS integrated intensity for components that originate from species on the $Ti_3C_2T_x$-surface. All O 1s XPS spectra were normalized at the background before background subtraction (thereof the arbitrary unit) and only relative amounts of O containing species can be obtained.

When the $Ti_3C_2T_x$ sample thereafter was exposed to $CO_2$ at 450 °C the amounts of O on the fcc-sites reduced significantly while the amount of O on the Ti-Ti bridge-sites increased to maximum intensity again. Similar as when H was present on the Ti on-top sites, the adsorption of $CO_2$ forced O on the fcc-sites to the Ti-Ti bridge-sites, although the main source must be the $H_2O$ diffusion from the bulk followed by dissociation. A relevant question is if $CO_2$ dissociates upon adsorption. There is a tiny structure at 287 eV in the C 1s XPS spectrum after $CO_2$ exposure that might indicate a very small amount of CO on the $Ti_3C_2T_x$ surface [8]. However, the information obtained in our study is not enough to resolve whether $CO_2$ may dissociate upon adsorption or not. Nevertheless, the low intensity in the 286-290 eV range of the C 1s XPS spectrum shows that the vast majority of the adsorbed $CO_2$ remains intact.

The third heat treatment removed the $CO_2$ and the O migrated back from the Ti-Ti bridge-sites to the fcc-sites. The third heat treatment was performed at 650 °C, which provided enough energy to proceed with the reaction between most of the remaining

H and O before desorption as indicated in Figure 8. With H and $CO_2$ removed the fcc-sites and Ti-Ti bridge-sites could be saturated with O and $H_2O$, respectively. $H_2O$ exposure did not influence the amount of the termination species on the $Ti_3C_2T_x$-surface.

### *3.6. Summary, remarks, and consequences of the findings*

The present study confirms that the termination species on as-prepared $Ti_3C_2T_x$, prepared through wet etching of $Ti_3AlC_2$, are F and O on the fcc-sites and on the Ti-Ti bridge-sites, respectively. In addition, a heat treatment significantly reduced the amount of F from the $Ti_3C_2T_x$-surface and the vacant fcc-sites were subsequently filled by O that migrated from the Ti-Ti bridge-sites. $H_2O$ was simultaneously diffusing from intercalation positions between the $Ti_3C_2T_x$-flakes to the newly available Ti-Ti bridge-sites. The process was further assisted by $H_2$ gas exposure at a moderate temperature increase that efficiently removed the remaining F leading to a saturated $Ti_3C_2T_x$-surface with O and $H_2O$ as termination species and H as adsorbate. The saturated $Ti_3C_2T_x$-surface was unable to capture $CO_2$. However, removing the H from the $Ti_3C_2T_x$-surface and the Ti on-top sites became available for $CO_2$ adsorption. Other adsorption sites for H and $CO_2$ adsorption cannot be excluded, although the O and $H_2O$ that were present as termination species may have limited the options in this study. Heat treatments up to 700 °C removed the H and $CO_2$ adsorbates but could not remove $H_2O$ from the $Ti_3C_2T_x$-surface. In fact, $H_2O$ is more strongly bonded to the $Ti_3C_2T_x$-surface than F. $H_2O$ can therefore be considered a termination species sitting on the Ti-Ti bridge-site. Furthermore, additional $H_2O$ exposure to a $Ti_3C_2T_x$-surface with termination species in a saturated condition shows water-splitting quality. Even so, while the H coverage increases because of $H_2O$ dissociation the amount of O-termination remained unchanged.

Consequently, $Ti_3C_2T_x$ as a $CO_2$ capturing material requires a preparation method that avoids intercalation of $H_2O$ between 2D $Ti_3C_2T_x$-flakes, because intercalated $H_2O$ between 2D $Ti_3C_2T_x$-flakes will become a source of $Ti_3C_2T_x$-surface poisoning for the $CO_2$ adsorption process. In addition, $Ti_3C_2T_x$ as a $CO_2$-capturing material requires a moisture-free environment. An additional implication with $H_2O$ as a termination species on the Ti-Ti bridge-site is that it will affect the intercalation of positive ions, such as $Li^+$ and $Na^+$. As discussed in a previous publication [20], O on the Ti-Ti bridge-site may act as anchor points in intercalation processes. The two electron lone pairs on the O are lobes with negative charge that are pointing out from the $Ti_3C_2T_x$-surface and will therefore attract positively charged species, such as positive ions and the positive end of dipolar molecules. The present study has shown that a $Ti_3C_2T_x$ sample can be intercalated with $H_2O$, which probably bonds to the O on the $Ti_3C_2T_x$-surface through hydrogen bonding. Hence, O in the Ti-Ti bridge-site could be a prerequisite for the intercalation process of $Li^+$ and/or $Na^+$, which probably have solvation spheres of $H_2O$ that bond to the O on the Ti-Ti bridge-sites through hydrogen bonding. When the O on the Ti-Ti bridge-sites are replaced by $H_2O$, with the hydrogen atoms pointing out from the $Ti_3C_2T_x$-surface, the $H_2O$ in the solvation spheres of positive ions will be repelled because the hydrogen atoms of the $H_2O$ in the solvation sphere will be pointing outwards. $H_2O$ on the Ti-Ti bridge-sites will, on the other hand, attract negative ions or the negative end of dipolar molecules. Hence, the present XPS study suggests that when $Ti_3C_2T_x$ is utilized as an ion storage material, the Ti-Ti bridge-site should be occupied by O for positive ions and $H_2O$ for negative ions.

One ion in particular that $Ti_3C_2T_x$ seems to be able to store is $H^+$ and that at room temperature and low pressure. The $H_2$ molecule dissociates on the $Ti_3C_2T_x$-surface and the adsorbed H lend its electron to the C layers in the $Ti_3C_2T_x$ crystal structure. The hydrogen can later on be released from the $Ti_3C_2T_x$ at a moderate temperature increase, restoring the $Ti_3C_2T_x$ to the same condition prior to the $H_2$ gas exposure (although the process requires an $H_2O$-free environment).

The present study has revealed a new termination species but also new information utilizing $Ti_3C_2T_x$ as ion storage material, $CO_2$ capture material, and a $H_2O$ splitting material. The obtained information demonstrates the requirements needed for particular application and, thus, guidance toward correct surface conditions when $Ti_3C_2T_x$ is activated in various future energy storage, $CO_2$ capture, or energy converting devices.

## 4. Conclusions

From the XPS study of the $Ti_3C_2T_x$ freestanding film we conclude that the as-prepared sample was almost completely saturated by F on the fcc-sites and almost completely saturated by O on the Ti-Ti bridge-sites, where most surface Ti that bond to F on the fcc-sites also bond to O on the Ti-Ti bridge-sites. Removing F from the $Ti_3C_2T_x$-surface through heat treatments and/or reaction with H enabled O-migration from the Ti-Ti bridge-sites to the fcc-sites that became vacant when the F were removed. The vacant Ti-Ti bridge-sites could then be occupied by $H_2O$ that have diffused from intercalated positions between $Ti_3C_2T_x$-flakes to the $Ti_3C_2T_x$-surface during the heat treatment. The $H_2O$ remained on the Ti-Ti bridge-sites even after several sample heat treatments at temperatures up to 700 °C. Hence, a new termination species is identified.

Hydrogen adsorbs on the $Ti_3C_2T_x$-surface and modification in the Ti 2p XPS spectrum indicates that the Ti on-top site on the $Ti_3C_2T_x$-surface is an H absorption site. The carbide peak in the C 1s XPS spectrum shows that the two C layers in the $Ti_3C_2T_x$ crystal structure gained charge when H was adsorbed.

With the $Ti_3C_2T_x$-surface completely covered by O, $H_2O$, and H there are no sites available for $CO_2$ adsorption. Some amount of $CO_2$ can, on the other hand, be adsorbed on the Ti on-top sites when the H is absent.

The adsorption of H and/or $CO_2$ on the Ti on-top site perturbs the molecular orbitals of the surface Ti enough to make the fcc-site less favorable for the O occupation. Some O will then migrate back to the Ti-Ti bridge-sites, which prevent adsorption of H and/or $CO_2$ on the Ti-Ti bridge-sites. Hence, to maximize the $CO_2$ adsorption, e.g., in an application as a carbon-capturing device, the $Ti_3C_2$-surface must be termination free and without the possibility for intercalated $H_2O$ to diffuse to the surface.

A $Ti_3C_2T_x$-surface, with O and $H_2O$ termination, has the ability to split water. Even so, there are no signs of adsorbed OH on the $Ti_3C_2T_x$-surface. This observation suggests desorption of OH-radicals or a complete dissociation followed by an $O_2$ formation and desorption.

**Acknowledgements**

We thank Dr Joseph Halim at Linköping University for preparing the sample and Dr Mikko-Heikki Mikkelä and Dr Margit Andersson for experimental support at the MAX IV Laboratory. Research conducted at MAX IV, a Swedish national user facility, is supported by the Swedish Research council under contract 2018-07152, the Swedish Governmental Agency for Innovation Systems under Contract 2018-04969, and Formas under Contract 2019-02496. We also thank the Swedish Government Strategic Research Area in Materials Science on Functional Materials at Linköping University (Faculty Grant SFO-Mat-LiU No. 2009 00971). M. M. acknowledges financial support from the Swedish Energy Research (Grant No. 43606-1) and the Carl Tryggers Foundation (CTS23:2746, CTS20:272, CTS16:303, CTS14:310).